\documentclass[11pt]{article}

\usepackage{graphicx}
\usepackage{hyperref}
\begin{document}
\title{Model-Based Iterative Reconstruction \\ for Radial Fast Spin-Echo MRI}
\author{Kai~Tobias~Block,~Martin~Uecker,~and~Jens~Frahm}


\date{}

\maketitle

\begin{abstract}
  In radial fast spin-echo MRI, a set of overlapping spokes with an
  inconsistent T2 weighting is acquired, which results in an averaged
  image contrast when employing conventional image reconstruction
  techniques. This work demonstrates that the problem may be overcome
  with the use of a dedicated reconstruction method that further allows
  for T2 quantification by extracting the embedded relaxation
  information.  Thus, the proposed reconstruction method directly yields
  a spin-density and relaxivity map from only a single radial data set.
  The method is based on an inverse formulation of the problem and
  involves a modeling of the received MRI signal. Because the solution
  is found by numerical optimization, the approach exploits all data
  acquired. Further, it handles multi-coil data and optionally allows
  for the incorporation of additional prior knowledge.  Simulations and
  experimental results for a phantom and human brain in vivo demonstrate
  that the method yields spin-density and relaxivity maps that are
  neither affected by the typical artifacts from TE mixing, nor by
  streaking artifacts from the incomplete k-space coverage at individual
  echo times.


\end{abstract}

\thispagestyle{empty}

\vfill
{\centering
\fbox{\parbox{\textwidth}{
\noindent
{\small
Copyright 2009 IEEE. Personal use of this material is permitted.
Permission from IEEE must be obtained for all other uses, in any current or future 
media, including reprinting/republishing this material for advertising or promotional 
purposes, creating new collective works, for resale or redistribution to servers or 
lists, or reuse of any copyrighted component of this work in other works.}

\vspace{0.2cm}
\noindent
Published in final form in: \\
IEEE Transactions on Medical Imaging 28:1759-1769 (2009) \\
DOI: \href{http://dx.doi.org/10.1109/TMI.2009.2023119}{10.1109/TMI.2009.2023119}
}}}

\enlargethispage{3\baselineskip}

\newpage      
\section{Introduction}

{F}{ast} spin-echo (FSE) MRI is one of the most frequently
used MRI techniques in today's clinical practice. In the FSE technique,
a train of multiple spin-echoes is generated after each RF excitation.
Therefore, it offers images with proton-density (PD) and T2 contrast at
a significantly reduced measuring time compared to a spin-echo sequence
with a single phase-encoding step per excitation. Although the k-space
lines that are sampled during each train of spin echoes have different
echo times (TE) and, thus, increasing T2 weighting, conventional
Cartesian sampling strategies still allow for a straightforward image
reconstruction with only minor artifacts under visual inspection when
grouping spin echoes with equal echo time around the center of k-space.
The underlying reason is the dominance of major image features by the
centrally located low spatial frequencies, while the periphery of
k-space rather defines edges which are less susceptible to changes in
the contrast weighting. Moreover, in the Cartesian case, the contrast of
the image can be adjusted by reordering the acquisition scheme such that
the central k-space lines are measured with the desired echo time. A
noticeable disadvantage of the T2 attenuation in k-space, however, is a
certain degree of image blurring. It arises because the signal decay can
be formally written as a multiplication of the true k-space data with
some envelope function, which yields a convolution in image space along
the phase encoding direction.

In recent years, radial sampling, as originally proposed by Lauterbur
\cite{Lau1973}, has regained considerable interest as an alternative to
Cartesian sampling. In this technique, coverage of k-space is
accomplished by sampling the data along coinciding spokes instead of
parallel lines. Radial sampling offers several promising advantages that
include a lower sensitivity to object motion \cite{Glo1992,Alt2002}, the
absence of any ghosting effects, and efficient undersampling properties.
In principle, radial trajectories can be combined with almost all MRI
techniques including FSE sequences \cite{Glo1992,Alt2002,Ras1999}.  In
this specific case, however, the modified sampling scheme has major
implications for the image reconstruction procedure. Because all spokes
pass through the center of k-space, each spoke carries an equal amount
of low spatial frequency information which, for different echo times,
exhibits pronounced differences in T2 contrast. This situation is
fundamentally different to the Cartesian case and poses complications
when employing conventional reconstruction methods such as filtered
backprojection or gridding \cite{Sul1985}: (i) The merging of spokes
with different echo times may cause streaking artifacts around areas
with pronounced T2 relaxation because the relative signal decay leads to
jumps in the corresponding point-spread-function \cite{The1999}. (ii)
The contrast of the image always represents an average of the varying T2
weightings. (iii) Differently ordered k-space acquisitions may no longer
be used to alternate the image contrast and, thus, to distinguish
between PD and T2 relaxation. On the other hand, because all spokes pass
through the center of k-space and capture the low spatial frequencies at
every echo time, a single radial FSE data set implicitly contains
information about the local signal decay. Therefore, dedicated
reconstruction methods have been developed that extract this embedded
temporal information in order to quantify the T2 relaxation within a
reduced measurement time relative to Cartesian-based T2 quantification
approaches.

Most of the existing techniques, such as k-space weighted image contrast
(KWIC) \cite{Son2000,Alt2005}, attempt to calculate a series of echo
time-resolved images by mixing the low spatial frequencies from spokes
measured at the desired echo time with high frequency information from
other spokes (measured at different echo times). A drawback of these
techniques is that the TE mixing tends to cause artifacts in the 
reconstructed images. Although their strength can be attenuated to some 
degree with specific echo mixing schemes \cite{Alt2005}, these artifacts 
limit the accuracy of the T2 estimates. 
In particular, the values become dependent on the object size as the change 
in k-space is assumed to be located only in the very central area that is 
densely covered by spokes measured at an equal echo time \cite{Alt2005}.

To overcome the limitation, this work demonstrates an iterative concept
for the reconstruction from radially sampled FSE acquisitions. Instead
of calculating any intermediate images, the proposed method estimates a
spin density and a relaxivity map directly from the acquired k-space
data using a numerical optimization technique. Because the procedure
involves a modeling of the received MRI signal that accounts for the
time dependency of the acquired samples, the approach exploits the
entire data for finding these maps without assuming that the contrast
changes are restricted to a very central area of k-space. Related ideas
have been presented by Graff et al \cite{Gra2006} and Olafsson et al
\cite{Ola2008}.

\section{Theory}

The method proposed here can be seen as an extension of a previous
development for the reconstruction from highly undersampled radial
acquisitions with multiple receive coils \cite{Blo2007}. In this work,
the reconstruction is achieved by iteratively estimating an image that,
on the one hand, is consistent with the measured data and, on the other
hand, complies with prior object knowledge to compensate for the missing
data.  For multi-echo data from a FSE sequence, this strategy is not
appropriate because it is impossible to find a single image that matches
the different contrasts at the same time. Therefore, it is necessary to
include the relaxation process into the signal modeling used to compare
the image estimate to the measured k-space samples. As the T2 relaxation
time is a locally varying quantity, this requires that the estimate
consists of a spin-density and a relaxation component instead of just an
intensity component, which directly yields a quantification of the
relaxation time. The objective of the present extended approach,
therefore, is to simultaneously find a spin-density map and a relaxivity
map such that snapshots, calculated for each echo time from these maps,
best match the spokes measured at the respective echo times.

\subsection{Cost Function}

In order to compute the maps, a cost function is needed which quantifies
the accuracy or quality of the match to the measured data, such that the
optimal solution can be identified by a minimum value of this function.
Because experimental MRI data is always contaminated by a certain degree
of Gaussian noise, the cost function of the proposed method uses the L2
norm and has the form
\begin{equation}
\label{eq:costfunc}
\Phi(\vec{\rho},\vec{r})=\frac{1}{2}\sum_t \sum_c \left\|\vec{F}(\vec{\rho},\vec{r},t,c) - \vec{y}_{t,c} \right\|^2_2 \;,
\end{equation} 
where $\vec{\rho}$ is a vector containing the values of the spin-density
map, and $\vec{r}$ is a vector containing values of the relaxivity map.
For a base resolution of $n \times n$ pixels, both vectors have $n^2$
entries. Further, $\vec{y}_{t,c}$ is a vector containing the raw data
from channel $c$ of all spokes measured at echo time $t$, where $c$ runs
from 1 to the total number of receive channels and $t$ runs over all
echo times. Finally, $\vec{F}$ is a vector function that calculates a
snapshot from the given spin-density and relaxivity map at echo time
$t$, and translates it to k-space using a Fourier transformation and
subsequent evaluation at the sampling positions of the spokes acquired
at time $t$. Moreover, before Fourier transformation, the mapping
includes a multiplication with the sensitivity profile of coil $c$.
These coil profiles are estimated from the same data set in a preceding
processing step using a sum-of-squares combination of the channels,
where it is assumed that the individual sensitivity profiles are smooth
functions.  A detailed description of this procedure is given in
\cite{Blo2007}.

The function $\vec{F}$ can be seen as the forward operation of the
reconstruction problem and comprises a model of the received MRI signal,
which is used to synthesize the corresponding k-space signal from the
given maps. The $j$th entry, i.e. the $j$th sample of the synthesized
data with k-space position $\vec{k}_j$ at echo time $t$, is given by
\begin{equation}
\label{eq:signalfunc}
F_j(\vec{\rho},\vec{r},t,c)= \sum_{\vec{x}\in \textrm{FOV}} \varrho(\vec{x}) \cdot e^{- R(\vec{x}) \cdot t} \cdot C_c(\vec{x}) \cdot e^{-i \, \vec{x} \cdot \vec{k}_j } \, ,
\end{equation}
where $\vec{x}$ denotes a position in image space such that the sum runs
over all (discrete) elements of the image matrix, and $C_c$ is the
complex sensitivity profile of the $c$th coil. Further, $\varrho$
denotes a function which evaluates the spin-density vector $\vec{\rho}$
at image position $\vec{x}$
\begin{equation}
\label{eq:densfunc}
\varrho(\vec{x})=\sum_i \rho_i \cdot \delta (\vec{x}-\vec{x_i}) \;,
\end{equation}
where $\rho_i$ denotes the $i$th component of the spin-density vector
with corresponding position $\vec{x_i}$ in image space. Accordingly, the
function $R$ evaluates the relaxivity vector $\vec{r}$
\begin{equation}
\label{eq:relaxfunc}
R(\vec{x})=\sum_i r_i \cdot \delta (\vec{x}-\vec{x_i}) \;.
\end{equation}
To reconstruct a set of measured data $\vec{y}$, a pair of vectors
$(\vec{\rho},\vec{r})$ has to be found that minimizes the cost function
$\Phi$. This can be achieved by using a numerical optimization method
that is suited for large-scale problems, like the non-linear
conjugate-gradient (CG) method.  Here, we employed the CG-Descent
algorithm recently presented by Hager and Zhang \cite{Hag2005}, which
can be used in a black box manner. Hence, it is only required to
evaluate $\Phi$ and its gradient at given positions
$(\vec{\rho},\vec{r})$ in the parameter space.

\subsection{Evaluation of Cost Function}

Evaluation of the cost function at a given pair $(\vec{\rho},\vec{r})$
can be done in a straightforward manner. However, to calculate the value
of $\Phi$ in a reasonable time, a practical strategy is to perform a
fast Fourier transformation (FFT) of the snapshots and to interpolate
the transforms onto the desired spoke positions in k-space using a
convolution with a radial Kaiser-Bessel kernel similar to the gridding
technique \cite{Sul1985,Jac1991}. Because the kernel is finite, it is
further necessary to pre-compensate for undesired intensity modulations
by multiplying the snapshots with an approximation of the kernel's
Fourier transform in front of the FFT -- commonly known as roll-off
correction \cite{Ras1999-2}.

The same strategy can be used for evaluating the gradient, which is a
vector containing the derivative of the cost function with respect to
each component of the spin-density and relaxivity vector. It is
convenient to decompose the problem into a separate derivation of $\Phi$
with respect to components of $\vec{\rho}$ and $\vec{r}$, respectively,
\begin{equation}
\label{eq:gradfunc}
\nabla \Phi= \left( \begin{array}{c}  \nabla_\rho \Phi \\ \nabla_r \Phi \end{array} \right)\;.
\end{equation}
To simplify the notation, the calculation is only shown for a single
time point and a single coil (indicated by $\phi$ instead of $\Phi$)
\begin{equation}
\label{eq:simplfunc}
\phi = \frac{1}{2}\left\|\vec{F} - \vec{y}\right\|^2_2\;.
\end{equation}
Derivation of Eq.~(\ref{eq:signalfunc}) with respect to components
of $\vec{\rho}$ gives 
\begin{equation}
\label{eq:fderiv}
\frac{\partial F_j}{\partial \rho_v}=e^{- R(\vec{x}_v) \cdot t} \cdot C_c(\vec{x}_v) \cdot e^{-i \, \vec{x}_v \cdot \vec{k}_j }\;,
\end{equation}
where $\vec{x}_v$ denotes the position of the $v$th component in image
space. Using the chain rule and inserting this equation then
yields (see Appendix) \begin{equation}
\label{eq:graddens}
\frac{\partial}{\partial \rho_v}\phi = e^{- R(\vec{x}_v) \cdot t} \cdot \Re \left\{  \overline{C_c(\vec{x}_v)} \cdot \sum_j \left( F_j - y_j \right) e^{i \, \vec{x}_v \cdot \vec{k}_j } \right\} .
\end{equation}
Hence, the gradient with respect to $\vec{\rho}$ can be obtained by
evaluating the cost function, calculating the residual, and performing
an inverse Fourier transformation, which is followed by a multiplication
with the complex conjugate of the coil profile and, finally, with the
relaxation term. Derivation of Eq.~(\ref{eq:signalfunc}) with
respect to components of $\vec{r}$ gives
\begin{equation}
\label{eq:derivrelax}
\frac{\partial F_j}{\partial r_v}= -t \cdot \varrho(\vec{x}_v) \cdot e^{- R(\vec{x}_v) \cdot t} \cdot C_c(\vec{x}_v) \cdot e^{-i \, \vec{x}_v \cdot \vec{k}_j } 
\end{equation}
and, in a similar way to (\ref{eq:graddens}), this yields
\begin{eqnarray}
\label{eq:gradrelax}
\frac{\partial}{\partial r_v}\phi & = & -t \cdot \varrho(\vec{x}_v) \cdot e^{- R(\vec{x}_v) \cdot t} \cdot \Re \left\{  
\overline{C_c(\vec{x}_v)} \cdot \sum_j \left( F_j - y_j \right) e^{i \, \vec{x}_v \cdot \vec{k}_j } \right\}.
\end{eqnarray}
Comparison with Eq.~(\ref{eq:graddens}) shows that the gradient with
respect to $\vec{r}$ can be easily obtained by multiplying the gradient
with respect to $\vec{\rho}$ with the components of the given
$\vec{\rho}$ and the echo time $t$. Of course, Eq.~(\ref{eq:graddens})
and Eq.~(\ref{eq:gradrelax}) have to be summed over each channel and
echo time occurring in the complete cost function~(\ref{eq:costfunc}).

\subsection{Regularization}

Because the aforementioned strategy for fast evaluation of the cost
function involves an interpolation step that is imperfect in practice,
and, further, because the measured values correspond to the continuous
Fourier transform whereas the estimated maps are discrete, it is
necessary to introduce a regularization of the estimate. If a
regularization is not used, implausible values arise after a certain
number of iterations in ill-conditioned areas of the estimate, i.e. in
entries of the maps that are not well-defined from the measured k-space
values.  For instance, the edges of the maps' Fourier transforms outside
of the sampled disc are ill-conditioned due to a lack of measured
samples in these areas. Because the optimizer simply tries to reduce the
total value of the cost function, it employs such ``degrees of freedom''
to minimize the residuum at neighboring sampling positions (it should be
noted that the interpolation strategy assumes that the true function can
be locally approximated by a weighted sum over the nearby grid
points). This results in local k-space peaks with high amplitude that 
translate into noise-alike image patterns for high iteration
numbers. The problem poses a general drawback of the iterative reconstruction
technique, which can be either adressed by stopping the optimization
procedure after some iterations or by regularizing the estimate to
suppress implausible values. Advantages of the latter approach are
that the estimate converges to a well-defined solution without being
sensitive to the total iteration number and that it allows for tuning
the solution by using different regularization techniques. A
regularization is accomplished by complementing the cost function with a
weighted penalty function $P(\vec{\rho},\vec{r})$ which can act on both
the spin-density and relaxivity map
\begin{eqnarray}
\label{eq:extcostfunc}
\Phi(\vec{\rho},\vec{r})&=&\frac{1}{2}\sum_t \sum_c \left\|\vec{F}(\vec{\rho},\vec{r},t,c) - \vec{y}_{t,c} \right\|^2_2 + \lambda \cdot P(\vec{\rho},\vec{r}) \;.
\end{eqnarray}

Different types of penalty functions can be used for the regularization.
For instance, constraining the total energy of the estimate by
penalizing the L2 norm of its components leads to the commonly known
(simple) Tikhonov regularization. In the present work, the L2 norms of
the finite differences of the maps' Fourier transforms are penalized,
which enforces certain smoothness of the k-space information 
\begin{eqnarray}
\label{eq:regterm}
P(\vec{\rho},\vec{r})=\left\| D_x \mathcal{F} \, \vec{\rho} \right\|^2_2
+ \left\| D_y \mathcal{F} \, \vec{\rho} \right\|^2_2
+ \left\| D_x \mathcal{F} \, \vec{r} \right\|^2_2
+ \left\| D_y \mathcal{F} \, \vec{r} \right\|^2_2 \;.
\end{eqnarray}
Here, $\mathcal{F}$ denotes the discrete Fourier transformation, and $D_x$ is the
finite difference operator in x-direction
\begin{equation}
D_x \, \textrm{I}(x,y)=\textrm{I}(x,y)-\textrm{I}(x-1,y)\;.
\end{equation}
It turned out that this choice offers robust suppression of local intensity accumulations 
in the ill-conditioned areas, so that no artifacts appear even for a high number of 
iterations (e.g., 1000 iterations). 

A drawback of any regularization technique is that it introduces a
weighting factor $\lambda$, which has to be chosen properly. If selected
too low, the regulariation becomes ineffective, whereas if selected too
high, the solution will be biased towards the minimization of the
penalty term (here leading to an undesired image modulation). A method for 
automatically determining the proper regularization weight would be highly 
desirable, but the establishment of such a method is yet an open issue.  
However, with the penalty term described in Eq.~(\ref{eq:regterm}), the reconstruction 
appears to be rather insensible to minor changes of the weighting factor $\lambda$.
In fact, all images presented here were obtained with a fixed value of
$\lambda=0.001$, which yielded an effective artifact suppression without
noticeably affecting the results. Presumably, the weight has to be adjusted for measured 
data with completely different scaling. In this case, a reasonable adjustment strategy 
is to start with a very small weight and to increase the value successively until the 
artifacts visually disappear.

\subsection{Initialization}

In the single-echo reconstruction scenario, it is very efficient to
initialize the optimizer with a properly scaled gridding solution.  This
choice significantly reduces the total number of iterations because the
optimizer starts with a reasonable guess. In the multi-echo case,
however, it is more difficult to obtain reasonable initial guesses for
the spin-density and relaxivity map, and several options exist. For
example, a curve fitting of either strongly undersampled or
low-resolution gridding solutions from single echo times could be used
to approximate the maps. Alternatively, an echo-sharing method like KWIC
could be employed. While preliminary analyses confirmed that these
strategies may lead to a certain acceleration of convergence, they also
indicated complications if the initial guesses contain implausible
values, for example, in relaxation maps which are obviously undefined in
areas with a void signal intensity. It is therefore necessary to remove
respective values from the initial guesses. The present work simply used
zero maps for the initialization, which require a higher number of
iterations but ensure a straightforward convergence to a reasonable
solution.

\subsection{Scaling and Snapshot Calculation}

Another factor with essential impact on the convergence rate is the
scaling of the time variable $t$. Although it intuitively makes sense to
directly use physical units, a proper rescaling of the time variable
significantly reduces the number of iterations.
Equation~(\ref{eq:gradrelax}) shows that the gradient with respect to
the relaxivity depends linearly on $t$, while this is not the case with
respect to the spin density. If the values of $t$ for the different
echoes are very small, then the cost function is much more sensitive to
changes in $\vec{\rho}$ and the problem is said to be \textit{poorly
  scaled} \cite{Noc2006}. In contrast, large values of $t$ lead to a
dominant sensitivity to perturbations in $\vec{r}$. Because finding a
reasonable solution requires a matching of both maps at the same time,
the optimization procedure is especially effective when the scaling of
$t$ is selected such that there is a balanced influence on the cost
function.  In our experience, a proper scaling, which depends on the
range of the object's spin-density and relaxivity values, allows for
reducing the number of required iterations from over 1000 iterations to
about only 80 iterations for a typical data set. Of course, a rescaling
of $t$ is accompanied by a corresponding scaling of the relaxivity
values in $\vec{r}$, which can be corrected afterwards to allow for
quantitative analyses. Noteworthy, the sensitivity to scaling is a
property of the specific optimization method used here (the non-linear
CG method) and not related to the reconstruction concept itself. Hence,
exchanging the CG method by an optimization technique that is scale
invariant (like the Newton's method) might render a rescaling
unnecessary but is accompanied by other disadvantages like calculation
of the Hessian \cite{Noc2006}.

Finally, after complete estimation of the spin-density map $\vec{\rho}$
and relaxivity map $\vec{r}$, snapshot images can be calculated for an
arbitrary echo time with
\begin{equation}
\label{eq:snapshot}
I_t(\vec{x})=\varrho(\vec{x}) \cdot e^{- R(\vec{x}) \cdot t}  \;.
\end{equation}
These images do not contain any additional information, but present the
estimated temporal information in a more familiar view.

\section{Methods}

\subsection{Data Acquisition}

For evaluation of the proposed reconstruction technique, simulated data
and experimental MRI data was used. The simulated data was created with
a numerical phantom which is composed of superimposed ellipses so that
the analytical Fourier transform of the phantom can be deduced from the
well-known Fourier transform of a circle. The numerical phantom mimics a
water phantom consisting of three compartments with different T2
relaxation times (200~ms, 100~ms, and 50~ms), which are surrounded by a
compartment with a relaxation time comparable to that of pure water
(1000~ms).  All experiments were conducted at 2.9~T (Siemens Magnetom
TIM Trio, Erlangen, Germany) using a receive only 12-channel head coil
in circularly polarized (CP) mode, yielding four channels with different
combinations of the coils. Measurements were performed for a water
phantom doped with $\textrm{MnCl}_2$ as well as the human brain in vivo,
where written informed consent was obtained in all cases prior to each
examination.

The simulated data and the experimental phantom data was acquired with a
base resolution of 160 pixels covering a FOV of 120~mm (bandwidth
568~Hz/pixel), while human brain data was acquired with a base
resolution of 224 pixels covering a FOV of 208~mm (bandwidth
360~Hz/pixel). A train of 16 spin echoes with an echo spacing of 10~ms
was recorded after a slice-selective 90$^\circ$ excitation pulse
(section thickness 3~mm). The spin echoes were refocused using a
conventional 180$^\circ$ RF pulse, enclosed by crusher gradients to
dephase spurious FID signals.  The total number of spokes per data set
ranged from 128 to 512 spokes, which were acquired using 8 to 32
excitations and measured with a repetition time of TR~=~7000~ms to avoid
saturation effects of the CSF. The "angular bisection" view-ordering
scheme was used as described by Song and Dougherty \cite{Son2000}, which
ensures that spokes measured at consecutive echo times have a maximum
angular distance. Noteworthy, this scheme is not required by the
proposed method, but it was employed to permit reconstructions with the
KWIC approach for comparison. Further, the sampling direction of every
second repetition was altered in such a way as to generate opposing
neighboring spokes. The procedure yields more tolerable artifacts in the
presence of off-resonances. Fat suppression was accomplished by a
preceding CHESS pulse, and an isotropic compensation mechanism was
applied to avoid gradient timing errors and corresponding smearing
artifacts due to misalignment of the data in k-space \cite{Spe2006}. For
comparison, a fully-sampled Cartesian data set of the human brain was
acquired with a multi-echo spin-echo sequence from the manufacturer
(base resolution 192~pixels, section thickness 4~mm, 16 echoes, echo
spacing 10ms, TR~=~7000~ms).  In this case, k-space was fully sampled at
all 16 echo times so that the acquisition time was about 22.4 minutes
for a single slice. Moreover, simulations with different degrees of
undersampling as well as simulations with added Gaussian noise were
performed for appraising the achievable reconstruction accuracy.

\subsection{Reconstruction}

All data processing was done offline using an in-house software package
written in C/C++. In a first step, phase offsets were removed by
aligning the phase of all spokes at the center of k-space. Coil
sensitivity profiles were estimated from the data set using the
procedure described in \cite{Blo2007}.  In addition, a thresholding mask
was obtained from the smoothed sum-of-squares image, so that areas with
void signal intensity could be set to zero by applying the mask to all
reconstructed images. For the interpolation in k-space from grid to
spokes and vice versa, a Kaiser-Bessel window with $L=6$,
$\beta=13.8551$ and twofold oversampling was used \cite{Bea2005}. To
speed up the iterations, the interpolation coefficients were
precalculated and stored in a look-up table.  The optimizer for
estimating the spin-density and relaxivity map was run for a fixed
number of 200 iterations to ensure that the estimate has converged,
although fewer iterations are usually sufficient for finding an accurate
solution.  Hence, an automatic stopping criterion would be highly
desirable for routine use to avoid unnecessary computation time. The
scaling of the time variable was chosen heuristically such that $t=300
\cdot n$ for the phantom study and $t=150 \cdot n$ for human brain data,
where $n$ is the echo number. Because this factor is object-dependent
and has significant impact on the optimization efficiency, it would be
highly desirable to employ an automatic mechanism for adjusting the
scaling. Such development, however, is outside the scope of this work.

For comparison, gridding reconstructions of the spokes measured at each
echo time were calculated using the same interpolation kernel.  Here,
the estimated coil sensitivity profiles were used to combine the
different channels instead of taking a sum-of-squares. Further,
time-resolved reconstructions employing the KWIC method were calculated.
In the initial KWIC work \cite{Son2000}, only 8 instead of 16 echoes were 
acquired per excitation. Therefore, we implemented two variants: either
high frequency information from all spokes was used to fill the outer
k-space area (kwic 16), or (in a sliding window manner) information from 
only the 8 neighboring echo times was shared (kwic 8). Apart form that, 
our implementation followed the basic KWIC approach described in \cite{Son2000}.
To allow for a fair comparison, the same
interpolation kernel was used, and coil profiles were employed for
channel combination.  Finally, spin-density and relaxivity maps were
estimated from the images by a pixelwise curve fitting using the
Levenberg-Marquardt algorithm.

\section{Results}

\subsection{Experimental Data}

\begin{figure}[t!]
\centering
\includegraphics[width=\textwidth]{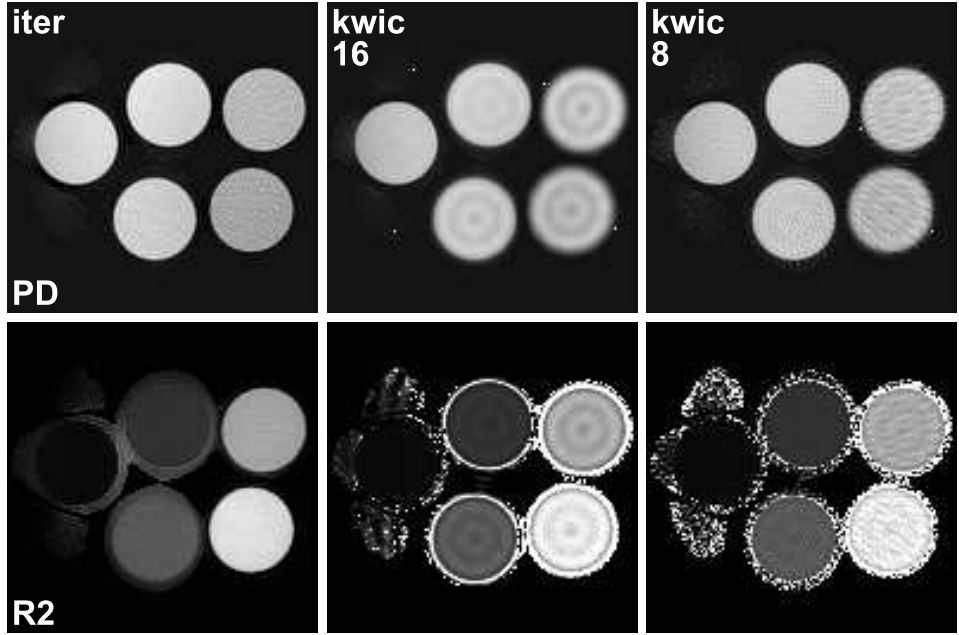}
\caption{
Spin-density maps (top) and relaxivity maps (bottom) estimated
for a phantom data set (base resolution 160~pixels, FOV 120~mm,
bandwidth 568~Hz/pixel bandwidth) using the proposed iterative method
(iter), KWIC combining all 16 echoes (kwic 16), and KWIC combining only
8 neighboring echoes (kwic 8). The images were acquired with a radial
FSE sequence using 32 repetitions and 16 echoes each, yielding a total
of 512 spokes. PD = proton density, R2 = T2 relaxivity.
}
\label{fig_phantom}
\end{figure}

Figure ~\ref{fig_phantom} compares spin-density and relaxivity maps for
a phantom containing five water-filled tubes with different
concentrations of $\textrm{MnCl}_2$, i.e. different T2 relaxation times,
which were estimated using the proposed method, the KWIC method sharing
all echoes, and the KWIC method sharing 8 neighboring echoes. It can be
seen that the sharing of k-space data in the KWIC reconstructions leads
to ring-like artifacts inside the tubes with fast T2 relaxation, in line
with the findings of Altbach et al \cite{Alt2005}. The artifacts are
more pronounced in the KWIC variant sharing all echoes, while the
variant sharing only 8 echoes suffers from streaking artifacts due to
incomplete coverage of the outer k-space. Such artifacts do not appear
in the iteratively estimated maps.  Here, the spin-density of the tube
with the shortest relaxation time is slightly underestimated, which is
probably caused by a higher amount of noise due to fast signal decay.
Further, because the relaxivity is undefined in areas with a void spin
density, the relaxivity maps are affected by spurious values outside of
the tubes in all cases. It should be noted that this effect is limited
to a narrow surrounding of the object due to the application of a
thresholding mask. In general, these spurious values are distinguishable 
from the object by a lack of intensity in either the spin-density map or 
the gridding image from all spokes.

\begin{figure}[t!]
\centering
\includegraphics[width=\textwidth]{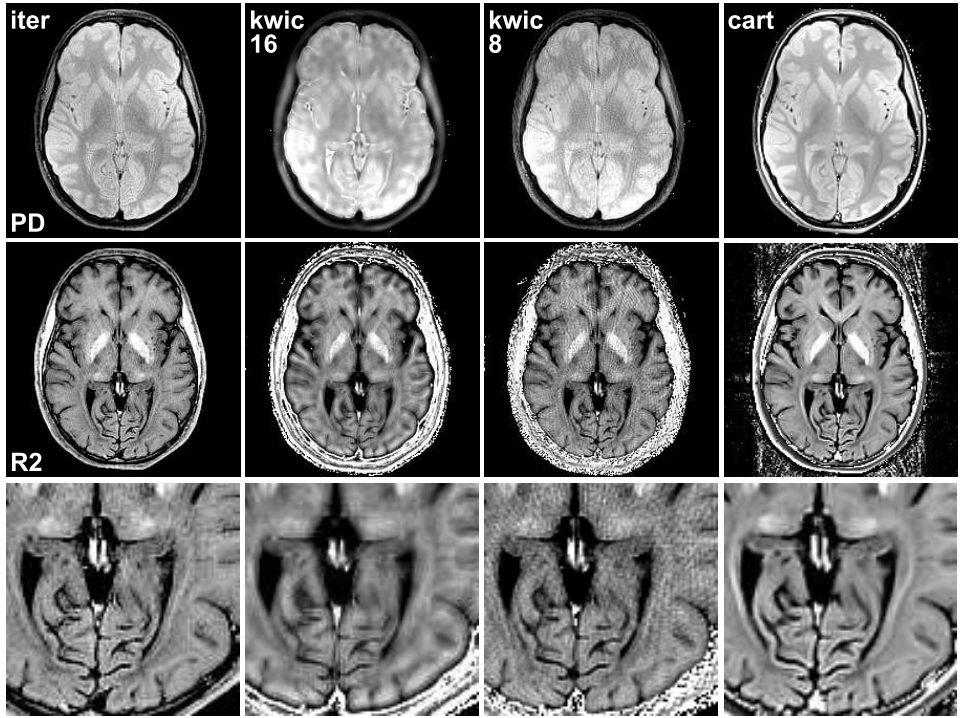}
\caption{
Spin-density maps (top) and relaxivity maps (bottom) estimated
for a transverse section of the human brain in vivo (base resolution
224~pixels, FOV 208~mm, bandwidth 360~Hz/pixel) using the proposed
method (iter), KWIC combining all 16 echoes (kwic 16), and KWIC
combining only 8 neighboring echoes (kwic 8). Other parameters as in
Fig.~\ref{fig_phantom}.  For comparison, maps from a fully-sampled
Cartesian reference data set are shown in the right column (cart). The
bottom row shows magnifications of the relaxivity maps.
}
\label{fig_brain}
\end{figure}

Figure \ref{fig_brain} shows corresponding reconstructions for a
transverse section of the human brain in vivo. Again, the KWIC
reconstruction using 8 echoes suffers from streaking artifacts, while
the maps involving all echoes appear fuzzy and blurry. In the latter case, 
the spin-density map is further contaminated by sharp hyperintense
structures. This results from padding the high frequencies 
with data from late echoes which introduces components with T2 weighting 
and poses a general problem when sharing data with varying contrast. 
The iteratively calculated maps
present without these artifacts. For comparison, maps from a
fully-sampled Cartesian data set are presented, which show good
agreement with the maps obtained from the iterative approach.
Noteworthy, because the slice thickness was higher in the Cartesian
acquisition, these maps show a slightly larger part of the frontal
ventricles, which, however, is not related to the reconstruction
technique.

\begin{figure}[t!]
\centering
\includegraphics[width=\textwidth]{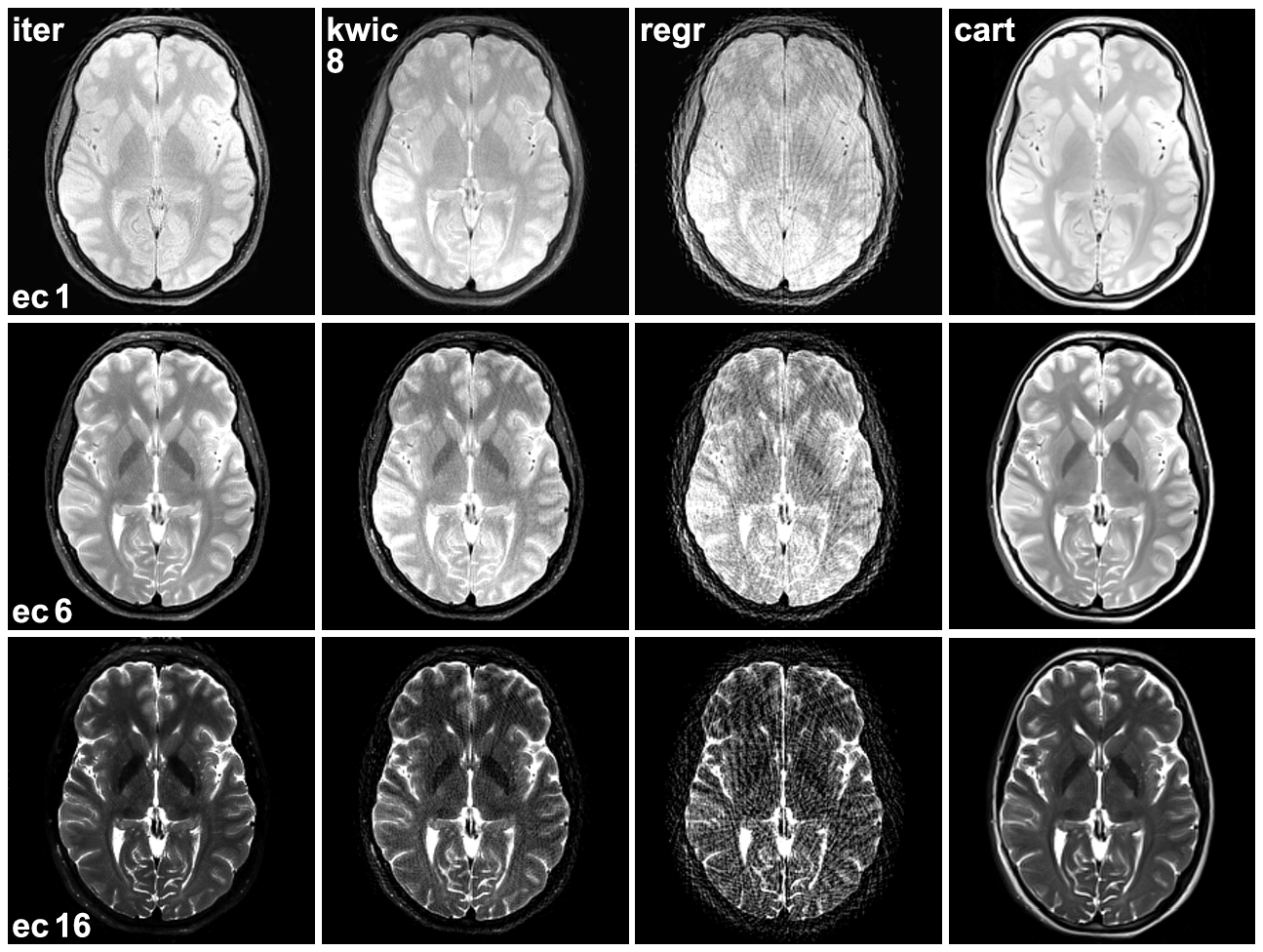}
\caption{
Snapshot reconstructions of the human brain (same data as
Fig.~\ref{fig_brain}) at the time of the first (ec1), 6th (ec6), and
last echo (ec16) using the proposed method (iter), KWIC combining 8
neighboring echoes (kwic 8), and direct gridding of the single echoes
(regr). The right column shows images from a fully-sampled Cartesian
data set (cart).
}
\label{fig_brainsnaps}
\end{figure}

For the same radial data set, Fig.~\ref{fig_brainsnaps} compares
snapshots of the first, 6th, and last echo reconstructed using the
proposed method with Eq.~(\ref{eq:snapshot}), direct gridding, and KWIC
with sharing of 8 echoes. Corresponding images from the Cartesian data
set are shown as reference. The contrast of the Cartesian and gridding
images can be taken as ground truth due to the equal echo time of all
k-space data used.  It can be seen that the snapshots calculated with
the proposed method show good match to the contrast of the Cartesian
gold standard while they are not affected by streaking artifacts.

\begin{figure}[t!]
\centering
\includegraphics{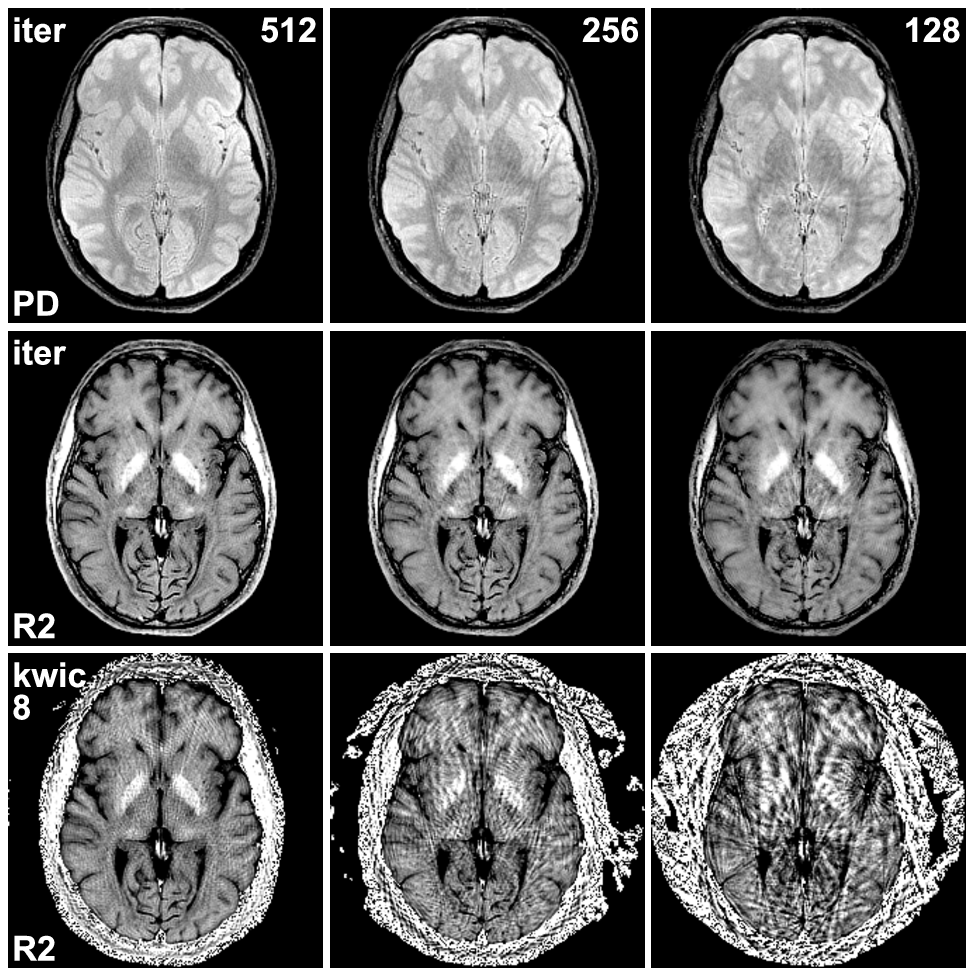}
\caption{
Spin-density maps (PD) and relaxivity maps (R2) estimated using the 
iterative approach for a transverse section of the human brain in vivo 
from (left) 512 spokes acquired with 32 repetitions, (middle) 256 spokes acquired 
with 16 repetitions, and (right) 128 spokes acquired with 8 repetitions 
(16 echoes each, base resolution 224~pixels, FOV 208~mm, bandwidth 360~Hz/pixel). 
The bottom row shows relaxivity maps obtained from the same data using the KWIC
approach (kwic8).
}
\label{fig_brainunders}
\end{figure}

Finally, Fig.~\ref{fig_brainunders} shows iterative reconstructions of
the human brain from radial data with different degrees of
undersampling, ranging from a total of 512 spokes (32 repetitions) to
only 128 spokes (8 repetitions). As expected, the data reduction is
accompanied by some loss of image quality, but even for 128 spokes
the iterative approach still offers a relatively good separation of
proton density and relaxivity. For comparison, relaxivity maps obtained
by the KWIC approach are shown in the bottom row, and it can be seen
that the image quality breaks down for higher degrees of undersampling.

\subsection{Simulated Data}

\begin{figure}[t!]
\centering
\includegraphics{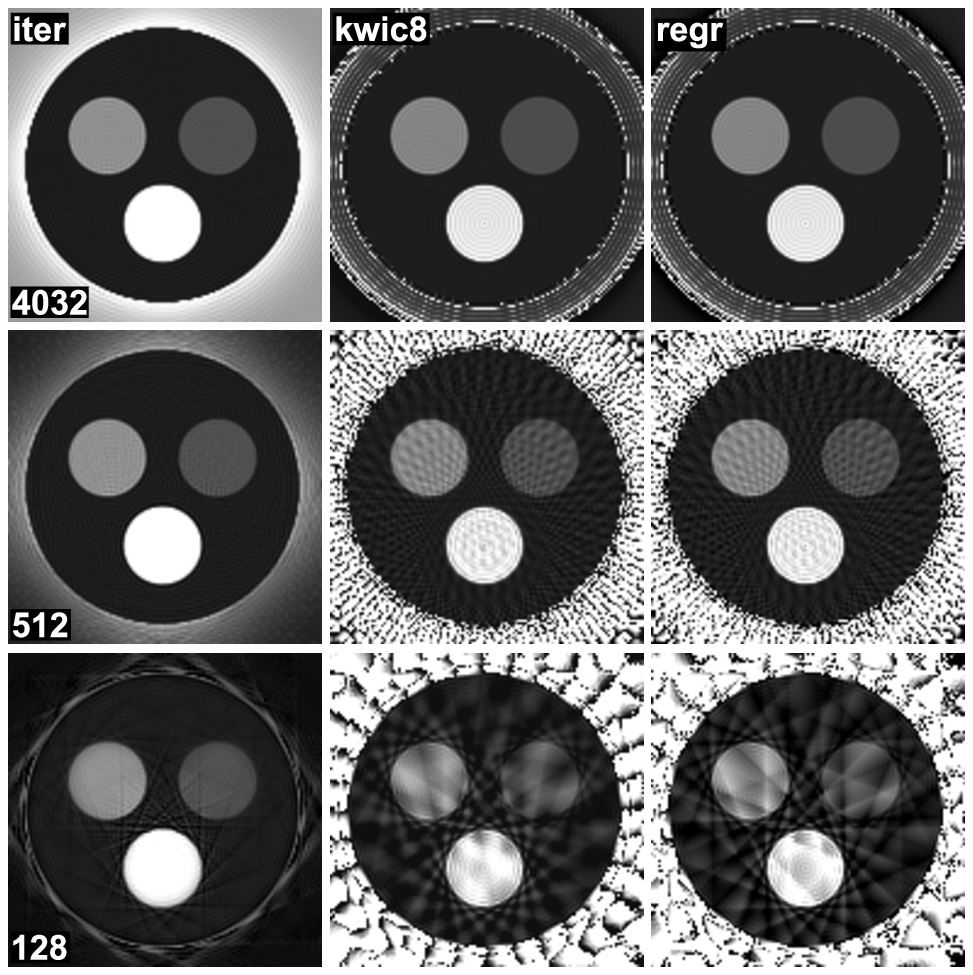}
\caption{
Relaxivity maps estimated from simulated data with (top) 4032 spokes 
from 252 repetitions (all echoes fully sampled), (middle) 512 spokes from 
32 repetitions, and (bottom) 128 spokes from 8 repetitions using (iter) 
the proposed iterative approach, (kwic8) the KWIC approach sharing 8 neigboring 
echoes, and (regr) direct gridding of the single echoes (16 echoes each, 
base resolution 160~pixels). The numerical phantom consists of compartments with
relaxations times of T2=200~ms, T2=100~ms, and T2=50~ms which are
surrounded by a compartment with T2=1000~ms.
}
\label{fig_accur}
\end{figure}

Figure \ref{fig_accur} shows relaxivity maps from simulated data
estimated with the proposed method, the KWIC method, and direct gridding
(the proton density maps are not shown as the proton density was set to
a constant value within the phantom). As in the prior figures, an
identical windowing function has been used for all relaxivity maps
within the figure, so that absolute relaxivity values were equally
mapped to grayscale values. The upper row corresponds to a fully sampled
data set (4032 spokes, 252 repetitions), and, hence, the number of spokes 
for each echo time complies
with the Nyquist theorem in the sense that the angular distance between
neighboring spokes is less or equal to $\Delta k=1/\textrm{FOV}$. In
this case, all three approaches yield maps without any streaking
artifacts. However, the maps created by KWIC and gridding present with
somewhat stronger Gibbs ringing artifacts, which are especially visible
within the smaller compartments. The middle row corresponds to the
degree of undersampling that was used in the experiments presented in
Fig.~\ref{fig_phantom}~--~Fig.~\ref{fig_brainsnaps} (512 spokes, 32 repetitions). 
Here, streaking artifacts appear for KWIC and gridding, whereas the iterative
reconstruction is free from these artifacts.  The bottom row shows maps
for a high degree of undersampling, corresponding to the highest
undersampling factor presented in Fig.~\ref{fig_brainunders} (128 spokes, 8 repetitions). 
For such an undersampling, minor streaking artifacts arise also in the iterative
reconstruction, while the KWIC and gridding reconstructions exhibit
severe streaking artifacts. Noteworthy, only a single receive channel
was generated in the simulations, and, thus, it clarifies that the
proposed method offers an improvement also without exploiting localized
coil sensitivities, which is implicitly done for experimental multi-coil
data due to the better conditioning of the problem. Table
\ref{tab_relaxivity} summarizes a region-of-interest (ROI) analysis of
the relaxivity maps from Fig.~\ref{fig_accur}, where identical regions
were analyzed in all maps. In all cases, the iterative approach
estimates the signal relaxivity with higher accuracy than gridding or
KWIC. Interestingly, even in the fully-sampled case, significant
deviations occur for the KWIC and gridding reconstruction. These
deviations result from the strong ringing effects that are apperent in
Fig.~\ref{fig_accur}, which appear to be more pronounced for radial
acquisitions than for Cartesian acquisitions. Hence, the strong signal
from the surrounding compartment smears into the quickly decaying
compartments, which causes a bias of the signal intensity in the
time-resolved images. The iterative approach seems to better cope with
this situation.

\begin{figure}[t!]
\centering
\includegraphics{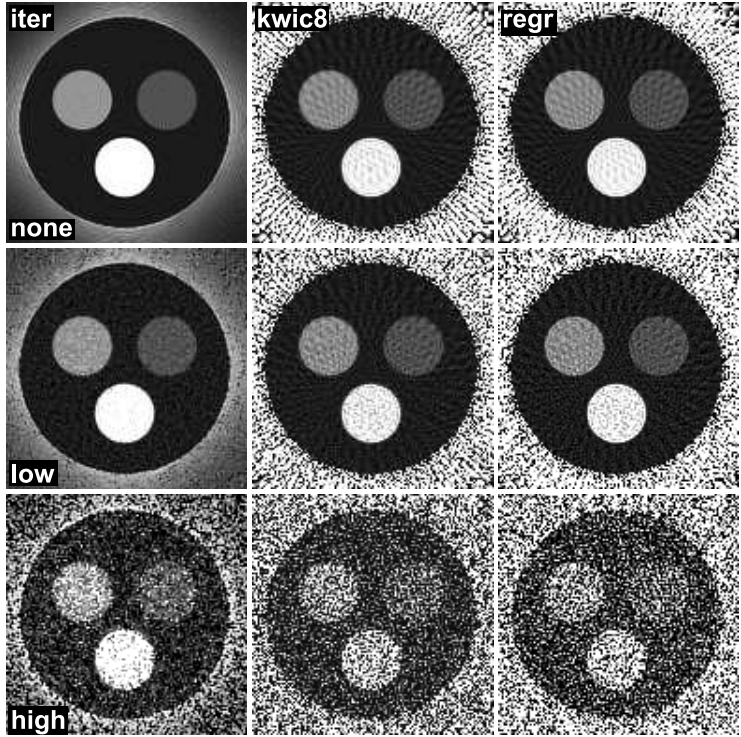}
\caption{
Relaxivity maps estimated from simulated data (top) without
noise, (middle) with a low level of Gaussian noise, and (bottom) with a
high level of Gaussian noise using (iter) the proposed iterative
approach, (kwic8) the KWIC approach sharing 8 neigboring echoes, and
(regr) direct gridding of the single echoes (512 spokes from 32 repetitions, 
16 echoes each, base resolution 160~pixels).  Relaxation times of the numerical
phantom as in Fig.~\ref{fig_accur}.
}
\label{fig_noise}
\end{figure}

\begin{table}[t!]
\makebox[\textwidth][c]{
                \begin{tabular}{ll|cccc}                
                \hline
                Noise level & & True T2 & Iterative & KWIC8 & Gridding \\
                \hline\hline
                \textit{none} & Compartment 1 & 50~ms & $50.2 \pm 0.1$~ms & $60.1 \pm 0.3$~ms & $59.8 \pm 0.3$~ms \\
                & Compartment 2 & 100~ms & $100.0 \pm 0.2$~ms & $110.3 \pm 0.7$~ms & $109.8 \pm 0.7$~ms \\
                & Compartment 3 & 200~ms & $199.9 \pm 0.6$~ms & $210.9 \pm 2.1$~ms & $210.5 \pm 2.2$~ms \\
                & Surrounding & 1000~ms & $996.5 \pm 11.9$~ms & $922.3 \pm 32.9$~ms & $930.4 \pm 37.8$~ms \\
    \hline
                \textit{low} & Compartment 1 & 50~ms & $50.0 \pm 0.3$~ms & $59.5 \pm 0.6$~ms & $58.5 \pm 0.8$~ms \\
                & Compartment 2 & 100~ms & $99.0 \pm 0.7$~ms & $109.7 \pm 1.1$~ms & $108.8 \pm 1.2$~ms \\
                & Compartment 3 & 200~ms & $200.5 \pm 1.8$~ms & $211.9 \pm 2.8$~ms & $210.9 \pm 3.2$~ms \\
                & Surrounding & 1000~ms & $1012.7 \pm 43.9$~ms & $901.8 \pm 43.3$~ms & $938.6 \pm 60.6$~ms \\
    \hline
                \textit{high} & Compartment 1 & 50~ms & $43.9 \pm 0.9$~ms & $48.6 \pm 4.2$~ms & $15.2 \pm 3.9$~ms \\
                & Compartment 2 & 100~ms & $76.6 \pm 2.3$~ms & $76.4 \pm 5.9$~ms & $28.9 \pm 10.2$~ms \\
                & Compartment 3 & 200~ms & $156.7 \pm 6.0$~ms & $155.2 \pm 10.0$~ms & $38.7 \pm 17.9$~ms \\
                & Surrounding & 1000~ms & $733.4 \pm 121.9$~ms & $257.9 \pm 73.5$~ms & $143.6 \pm 228.4$~ms \\
    \hline
                \end{tabular}
}
        \caption{ROI analysis of the relaxivity maps shown in Fig.~\ref{fig_noise}.}
        \label{tab_noise}
\end{table}

Finally, Fig.~\ref{fig_noise} shows relaxivity maps estimated from
simulated data with different degrees of added Gaussian noise (512 spokes, 32
repetitions). It demonstrates that the proposed approach
works stable also under noisy conditions, which is somewhat obvious as
the matching of the maps to the measured data is done in a least-squares
way. The maps estimated with KWIC and gridding present with a slightly
higher noise level. The results of a ROI analysis of the maps are
summarized in Table \ref{tab_noise}.

\section{Discussion}

As demonstrated, existing reconstruction methods for radial FSE that
share data from different echo times always require a trade-off between
the accuracy of the image contrast and an undersampling of the outer
k-space, which results in streaking artifacts. The main advantage of the
proposed method is that the use of a signal model allows for the
combination of spokes measured at different times. Thus, it exploits all
sampled data without having to assume that contrast changes are limited
to some central part of k-space.

\subsection{Computational Load}

A disadvantage of the present implementation, however,
is a significantly higher computational requirement than for
conventional non-iterative methods.  Because an individual Fourier
transformation with subsequent gridding is required for each echo time
and receiver channel, a single evaluation of the cost function for the
examples analyzed here involves 64 FFT and gridding steps. 
Evaluating the gradient requires even twice the number of operations,
 and one full iteration of the algorithm often needs several
evaluations of the cost function and gradient.

However, many of the operations can be performed in parallel. In the
proof-of-principle implementation, we used the OpenMP interface to
parallelize the calculations for different echo times. Hence, the
evaluation of the cost function and gradient is executed on different
cores at the same time. Using a system equipped with two Intel Xeon
E5345 quad core processors running at 2.33 GHz, the 200 iterations took
about two minutes per slice (excluding the calculation of the look-up
table, which takes additional 8 seconds on a single core). Despite
foreseeable progress in multi-core processor technology which promises
significant acceleration, a use of the method in near future is likely
to be limited to applications where delayed reconstructions are tolerable. 
However, because preliminary gridding reconstructions could be calculated 
to provide immediate feedback to the operator, this limitation might be 
secondary in clinical practice.

\subsection{Accuracy}

\begin{table}[t!]
\makebox[\textwidth][c]{
                \begin{tabular}{ll|cccc}                
                \hline
                Data set & & True T2 & Iterative & KWIC8 & Gridding \\
                \hline\hline
                \textit{4032 spokes} & Compartment 1 & 50~ms & $49.9 \pm 0.1$~ms & $59.7 \pm 0.2$~ms & $59.8 \pm 0.2$~ms \\
                & Compartment 2 & 100~ms & $100.0 \pm 0.1$~ms & $109.9 \pm 0.2$~ms & $109.9 \pm 0.2$~ms \\
                & Compartment 3 & 200~ms & $199.9 \pm 0.4$~ms & $209.9 \pm 0.2$~ms & $209.9 \pm 0.2$~ms \\
                & Surrounding & 1000~ms & $1001.0 \pm 4.7$~ms & $930.6 \pm 1.0$~ms & $930.6 \pm 1.0$~ms \\
    \hline
                \textit{512 spokes} & Compartment 1 & 50~ms & $50.2 \pm 0.1$~ms & $60.1 \pm 0.3$~ms & $59.8 \pm 0.3$~ms \\
                & Compartment 2 & 100~ms & $100.0 \pm 0.2$~ms & $110.3 \pm 0.7$~ms & $109.8 \pm 0.7$~ms \\
                & Compartment 3 & 200~ms & $199.9 \pm 0.6$~ms & $210.9 \pm 2.1$~ms & $210.5 \pm 2.2$~ms \\
                & Surrounding & 1000~ms & $996.5 \pm 11.9$~ms & $922.3 \pm 32.9$~ms & $930.4 \pm 37.8$~ms \\
    \hline
                \textit{128 spokes} & Compartment 1 & 50~ms & $49.1 \pm 0.1$~ms & $57.6 \pm 0.3$~ms & $60.0 \pm 0.5$~ms \\
                & Compartment 2 & 100~ms & $98.8 \pm 0.2$~ms & $105.7 \pm 1.1$~ms & $108.9 \pm 1.3$~ms \\
                & Compartment 3 & 200~ms & $197.1 \pm 0.7$~ms & $205.4 \pm 3.5$~ms & $213.7 \pm 4.1$~ms \\
                & Surrounding & 1000~ms & $1032.3 \pm 14.0$~ms & $901.7 \pm 41.5$~ms & $1073.8 \pm 106.5$~ms \\
    \hline
                \end{tabular}
}
        \caption{ROI analysis of the relaxivity maps shown in Fig.~\ref{fig_accur}.}
        \label{tab_relaxivity}
\end{table}

From a theoretical point of view, the proposed method should make
optimal use of all data measured and, thus, deliver a high accuracy,
which is confirmed by the results listed in Table \ref{tab_relaxivity}
for a simulated phantom. Because the solution is found in a least-squares 
sense, this should also hold true for data contaminated by noise as verified
in Fig.~\ref{fig_noise}. In practice, however, there are a number of 
factors that might affect the achievable experimental accuracy.

First, the procedure used to determine the coil sensitivities is simple
and might introduce a bias due to inappropriate characterization of the
profiles. In particular, the procedure fails in areas with no or very
low signal intensity, so that routine applications will probably require
a more sophisticated procedure. Second, the Fourier transform of the
object, as encoded by the MRI signal, is non-compact. Therefore, any
finite sampling is incomplete, which makes it impossible to invert the
spatial encoding exactly. Consequently, truncation artifacts arise when
employing a discrete Fourier transformation (DFT), which present as a
convolution of the object with a sinc function. Because DFTs are used to
compare the snapshots to the measured data and, further, because the
truncation artifacts are different for each echo time, this effect might
interfere with the estimation of a solution that is fully consistent
with all measured data. In particular, ringing patterns around
high-intensity spots might lead to a bias of surrounding pixels that
possibly diverts the decay estimated in these areas.  Noteworthy, this
effect is an inherent problem of any MRI technique and not limited to
the proposed method, as evident from the deviations occuring in the
gridding reconstructions in Table \ref{tab_relaxivity}.

Finally, if the relaxation process is so fast that the signal decay is
insufficiently captured by the acquired echo train, inaccurate
spin-density and relaxivity values will be estimated. For example, if a
signal intensity above noise level is received only at the first echo
time, the algorithm will probably assume a too low spin-density and a
too low relaxivity, which would likewise describe the observed signal
intensities in a least-squares sense. However, this is a general problem
of any T2 estimation technique and can only be overcome by a finer
temporal sampling. Also, inaccuracies that might occur when the actual
relaxation process differs from a pure mono-exponential decay are not
limited to the present method.  In fact, any T2 estimation technique has
to employ a simplified signal model at some stage for quantifying the
relaxation and, consequently, deviations from an assumed exponential
signal decay always impair the estimation accuracy.

\subsection{Extensions}

Although focused on the reconstruction of FSE data, the method can be
used for multi-echo data from other sequences as well. Depending on the
contrast mechanism of the individual sequence, it might be necessary to
adapt the signal model (\ref{eq:signalfunc}). Further, for non-refocused
multi-echo sequences the data can be significantly affected by
off-resonance effects due to the pronounced sensitivity of radial
trajectories. In this case, it might be possible to map also the
off-resonances by replacing the relaxivity with a complex-valued
parameter and adjusting the gradient of the cost function. However, due
to the extended parameter space it is expected that this strategy will
be only successful if suitable constraints for the estimates are
incorporated. This can be achieved by extending the cost function
Eq.~(\ref{eq:extcostfunc}) by additional penalty terms that imply
certain prior knowledge about the solution. For example, if it can be
assumed that the object is piecewise-constant to some degree or 
has a sparse transform representation, it is reasonable to penalize the total 
variation of the maps \cite{Blo2007} or to apply a constraint on the transform
representation \cite{Lus2007}. Further, it might be beneficial to penalize relaxation times that
obviously exceed the range of plausible values, e.g negative relaxation
times or relaxation times greater than the repetition time TR.

Moreover, the reconstruction concept is not only applicable to data with
different contrast due to spin relaxation or saturation, but can be
adapted to completely different imaging situations as well. In this
regard, the current work demonstrates the feasibility of extending the
inverse reconstruction scheme to more complex imaging problems that
require a non-linear processing. A prerequisite for the application to
other problems, however, is that a simple analytical signal model,
comparable to Eq.~(\ref{eq:signalfunc}), can be formulated.  Further, it
is required that the derivative of the signal model with respect to all
components of the parameter space can be calculated, and that the model
allows for a relatively fast evaluation of the cost function and its
gradient.

\section{Conclusion}

This work presents a new concept for iterative reconstruction from
radial multi-echo data with a main focus on fast spin-echo acquisitions.
Instead of sharing k-space data with different echo times to approximate
time-resolved images, the proposed method employs a signal model to
account for the time dependency of the data and directly estimates a
spin-density and relaxivity map. Because the approach involves a
numerical optimization for finding a solution, it exploits all data
sampled and allows for an efficient T2 quantification from a single
radial data set. In comparison with Cartesian quantification techniques,
such data can be acquired in a shorter time and with less motion
sensitivity. The method is computationally intensive and presently
limited to applications where a delayed reconstruction is acceptable.

\newpage
\section*{Appendix}

The (simplified) cost function $\phi$ defined in
Eq.~(\ref{eq:simplfunc}) can be written as
\begin{eqnarray}
\phi &=& \frac{1}{2}\left\|\vec{F} - \vec{y}\right\|^2_2 
=\frac{1}{2}\sum_j \left( F_j - y_j \right) \overline{\left( F_j - y_j \right)} \nonumber \\
&=& \frac{1}{2}\sum_j F_j \overline{F_j} + y_j \overline{y_j} - y_j \overline{F_j} - \overline{y_j} F_j\;,
\end{eqnarray}
where $\overline{(\ \cdot\ )}$ denotes the complex conjugate.
The derivative of this function with respect to any component $u$ of the estimate 
vector is obtained using the chain rule
\begin{eqnarray}
\frac{\partial}{\partial u}\phi &=& 
\frac{1}{2}\sum_j F_j \frac{\partial}{\partial u} \overline{F_j}+
\overline{F_j} \frac{\partial}{\partial u} F_j - \overline{y_j} \frac{\partial}{\partial u} F_j - y_j \frac{\partial}{\partial u} \overline{F_j} \nonumber \\
&=& \frac{1}{2}\sum_j \left( F_j - y_j \right) \frac{\partial}{\partial u} \overline{F_j}+ \left( \overline{F_j} - \overline{y_j} \right) \frac{\partial}{\partial u} F_j \nonumber \\
&=& \frac{1}{2}\sum_j \left( F_j - y_j \right) \frac{\partial}{\partial u} \overline{F_j}+ \frac{1}{2} \overline{\sum_j \left( F_j - y_j \right) \frac{\partial}{\partial u} \overline{F_j}} \nonumber \\
&=& \Re \left\{ \sum_j \left( F_j - y_j \right) \frac{\partial}{\partial u} \overline{F_j} \right\}\;.
\end{eqnarray}
Inserting Eq.~(\ref{eq:fderiv}) then yields the derivative of the cost
function with respect to a component of the spin-density map $\rho_v$
\begin{eqnarray}
\frac{\partial}{\partial \rho_v}\phi &=&
\Re \left\{ \sum_j \left( F_j - y_j \right) \frac{\partial}{\partial \rho_v} \overline{F_j} \right\} \nonumber \\
&=& \Re \left\{ \sum_j \left( F_j - y_j \right) e^{- R(\vec{x}_v) \cdot t} \cdot \overline{C_c(\vec{x}_v)} \cdot e^{i \, \vec{x}_v \cdot \vec{k}_j } \right\} \nonumber \\
&=& e^{- R(\vec{x}_v) \cdot t} \cdot \Re \left\{  \overline{C_c(\vec{x}_v)} \cdot \sum_j \left( F_j - y_j \right) e^{i \, \vec{x}_v \cdot \vec{k}_j } \right\}\;.
\end{eqnarray}
The derivative with respect to components of the relaxivity map can be 
obtained accordingly.

\newpage



\begin{thebibliography}{1}
  
\bibitem{Lau1973} P.D.~Lauterbur, ``Image formation by induced local
  interactions: Examples employing nuclear magnetic resonance'',
  \emph{Nature}, vol.~242, pp.~190--191, 1973.

\bibitem{Glo1992} G.H.~Glover and J.M.~Pauly, ``Projection
  reconstruction techniques for reduction of motion effects in MRI'',
  \emph{Magn Reson Med}, vol.~28, pp.~275--289, 1992.
  
\bibitem{Alt2002} M.I.~Altbach, E.K.~Outwater, T.P.~Trouard,
  E.A.~Krupinski, R.J.~Theilmann, A.T.~Stopeck, M.~Kono, and
  A.F.~Gmitro, ``Radial fast spin-echo method for T2-weighted imaging
  and T2 mapping of the liver'', \emph{J Magn Reson}, vol.~16,
  pp.~179--189, 2002. 

\bibitem{Ras1999} V.~Rasche, D.~Holz, and W.~Schepper, ``Radial Turbo
  Spin Echo Imaging'', \emph{Magn Reson Med}, vol.~32, pp.~629--638,
  1994.  
  
\bibitem{Sul1985} J.D.~O'Sullivan, ``A fast sinc function gridding
  algorithm for {Fourier} inversion in computer tomography'', \emph{IEEE
    Trans on Med Imaging}, vol.~4, pp.~200--207, 1985.  
  
\bibitem{The1999} R.J.~Theilmann, A.F.~Gmitro, M.I.~Altbach, and T.P.~Trouard, 
  ``View-ordering in radial fast spin-echo imaging'', \emph{Magn Reson Med}, 
  vol.~51, pp.~768--774, 2004.
  
\bibitem{Son2000} H.K.~Song and L.~Dougherty, ``k-Space weighted
  image contrast (KWIC) for contrast manipulation in projection
  reconstruction MRI'', \emph{Magn Reson Med}, vol.~44, pp.~825--832, 2000.
  
\bibitem{Alt2005} M.I.~Altbach, A.~Bilgin, Z.~Li, E.W.~Clarkson, T.P.~Trouard,
  and A.F.~Gmitro, ``Processing of radial fast spin-echo data for
  obtaining T2 estimates from a single k-space data set'', \emph{Magn
  Reson Med}, vol.~54, pp.~549--559, 2005.
  
\bibitem{Gra2006} C.~Graff, Z.~Li, A.~Bilgin, M.I.~Altbach, A.F.~Gmitro
  and E.W.~Clarkson, ``Iterative T2 Estimation from Highly Undersampled
  Radial Fast Spin-Echo Data'', \emph{Proc Intl Soc Mag Reson Med},
  vol.~14, p.~925, 2006.
 
\bibitem{Ola2008} Olafsson, V.T. and Noll, D.C. and Fessler, J.A.,
  ``Fast Joint Reconstruction of Dynamic $R_2^*$ and Field Maps in
  Functional MRI'', \emph{IEEE Trans on Med Imaging}, vol.~27,
  pp.~1177--1188, 2008.
  
\bibitem{Blo2007} K.T.~Block, M.~Uecker, and J.Frahm, ``Undersampled
  radial MRI with multiple coils. Iterative image reconstruction using a
  total variation constraint'', \emph{Magn Reson Med}, vol.~57,
  pp.~1086-1098, 2007.
     
\bibitem{Hag2005} W.W.~Hager and H.~Zhang, ``A new conjugate gradient
  method with guaranteed descent and an efficient line search'',
  \emph{SIAM J Optimization}, vol.~16, pp.~170--192, 2005.
  
\bibitem{Jac1991} J.~Jackson, C.H.~Meyer, D.G.~Nishimura, and
  A.~Macovski, ``Selection of a convolution function for Fourier
  inversion using gridding'', \emph{IEEE Trans on Med Imaging}, vol.~10,
  pp.~473--478, 1991.
  
\bibitem{Ras1999-2} V.~Rasche, R.~Proksa, P.~B\"ornert, and H.~Eggers,
  ``Resampling of data between arbitrary grids using convolution
  interpolation'', \emph{IEEE Trans on Med Imaging}, vol.~18,
  pp.~385--392, 1999.
  
\bibitem{Noc2006} J.~Nocedal and S.J.~Wright, \emph{Numerical
    Optimization}, Springer, 2006.
  
\bibitem{Spe2006} P.~Speier and F.~Trautwein, ``Robust radial imaging
  with predetermined isotropic gradient delay correction'', \emph{Proc
    Intl Soc Mag Reson Med}, vol.~14, p.~2379, 2006.
  
\bibitem{Bea2005} P.J.~Beatty, D.G.~Nishimura, and J.M.~Pauly, ``Rapid
  gridding reconstruction with a minimal oversampling ratio'',
  \emph{IEEE Trans on Med Imaging}, vol.~24, pp.~799--808, 2005.

\bibitem{Lus2007} M.~Lustig, D.~Donoho, and J.M.~Pauly, ``Sparse MRI: The application 
  of compressed sensing for rapid MR imaging'', \emph{Magn Reson Med}, vol.~58,
  pp.~1182-1195, 2007.

\end{thebibliography}
\end{document}